\documentstyle[preprint,aps]{revtex}
\begin{document}
\input{epsf}
\draft
\newfont{\form}{cmss10}
\newcommand{\e}{\varepsilon}
\renewcommand{\b}{\beta}
\newcommand{\unity}{1\kern-.65mm \mbox{\form l}}
\newcommand{\D}{D \raise0.5mm\hbox{\kern-2.0mm /}}
\newcommand{\A}{A \raise0.5mm\hbox{\kern-1.8mm /}}
\def\pmb#1{\leavevmode\setbox0=3D\hbox{$#1$}\kern-.025em\copy0\kern-\wd0
\kern-.05em\copy0\kern-\wd0\kern-.025em\raise.0433em\box0}
\def\D{\hbox{\hbox{${D}$}}\kern-1.9mm{\hbox{${/}$}}}
\def\kbar{\hbox{$k$}\kern-0.2truecm\hbox{$/$}}
\def\nbar{\hbox{$n$}\kern-0.23truecm\hbox{$/$}}
\def\pbar{\hbox{$p$}\kern-0.18truecm\hbox{$/$}}
\def\nhbar{\hbox{$\hat n$}\kern-0.23truecm\hbox{$/$}}
\newcommand{\dif}{\hspace{-1mm}{\rm d}}
\newcommand{\dil}[1]{{\rm Li}_2\left(#1\right)}
\newcommand{\diff}{{\rm d}}
\title{The instanton contributions to Yang-Mills theory on the torus: 
localization, Wilson loops and the perturbative expansion}
\author{L. Griguolo }
\address{Dipartimento di Fisica "G. Galilei",
INFN, Sezione di Padova,\\
Via Marzolo 8, 35131 Padua, Italy}
\maketitle
\begin{abstract}
The instanton contributions to the partition function and to 
homologically trivial Wilson loops for a $U(N)$ Yang-Mills theory on a 
torus $T^2$ are analyzed. An exact expression for the partition function 
is obtained as a sum of contributions localized around the classical 
solutions of Yang-Mills equations, that appear according to the 
general classification of Atiyah and Bott. Explicit expressions for the 
exact Wilson loop averages are obtained when $N=2$, $N=3$. For general 
$N$ the contribution of the zero-instanton sector has been carefully 
derived in the decompactification limit, reproducing the sum of the 
perturbative series on the plane, in which the light-cone gauge Yang-Mills 
propagator is prescribed according to Wu-Mandelstam-Leibbrandt (WML). 
Agreement with the results coming from $S^2$ is therefore obtained, 
confirming the truly perturbative nature of the WML computations.
\end{abstract}
\vskip 2.0truecm
Padova preprint DFPD 98/TH .


\vskip 3.0truecm
\vfill\eject

\narrowtext
\section{Introduction}
\noindent
In the last few years a lot of attention has been devoted to quantum 
Yang-Mills theories on compact two-dimensional surfaces: 
in spite of their seeming triviality, many interesting (and highly 
non-trivial) results were obtained exploiting their non-perturbative 
solvability. The exact partition function was in fact 
derived long ago by Migdal \cite{Mig}, using a lattice formulation 
of the theory, and subsequently the same technique was extended to the 
computation of Wilson loop averages \cite{Ruso}. Starting from these 
results it has been recognized \cite{Wati} that, in the large-$N$ limit, 
two-dimensional $U(N)$ Yang-Mills theory is a string theory, a fact that 
is widely believed to have a four-dimensional counterpart. 

On the other hand, from a conventional quantum field theory point of view, 
${\rm YM}_{2}$ exhibits an interesting and peculiar behaviour, that is 
actually at the basis of its exact solvability: the partition function is 
given by a sum over contributions localized at classical solutions of 
the theory. For finite ${N}$ the path-integral is represented as sum 
over unstable instantons, where each instanton contribution is 
modified by a finite, but non-trivial, perturbative expansion that 
carries the quantum corrections. Witten has shown \cite{Witte} how to 
understand, from a general point of view, this behaviour constructing a 
non abelian generalization of the Duistermaat-Heckman theorem 
\cite{mati} and exploiting the relation with a topological field theory.
  
A further intriguing aspect is the appearance, on genus zero and
in the limit of large $N$, of a third order phase transition at some critical
value of the sphere area \cite{Kaza}: a strong coupling phase, 
where a pure area exponentiation for Wilson loops  dominates in the 
decompactification limit, is distinguished from a weak coupling phase with no
confining behaviour \cite{Lupi}. Again instantons provide a clear physical 
picture of this phenomenon: the explicit computation of instanton 
contributions performed in \cite{Grosso} has shown that the deconfining 
phase comes as a result of instantons condensation. Topologically 
non-trivial configurations play an analogous role on the cylinder, as 
noticed some time before in \cite{Dadda}, leading to a similar phase transition.

Rather surprisingly, it has been recently realized \cite{Us} that the 
instanton picture is essential in understanding  some controversial 
aspects of ${\rm YM}_2$ (and ${\rm QCD}_2$) on the plane, some of which 
dating back to the mid-seventies. The facts are the following. 
Confinement in ${\rm QCD_2}$ is usually regarded as a perturbative feature. 
As a matter of fact the area exponentation is simply obtained (for any $N$) 
by summing on the plane the conventional perturbative series in 
light-cone gauge (where the situation is trivialized by the absence of 
self-interaction terms), with the 't Hooft-CPV prescription for the gluon 
exchange potential \cite{Hooft}. The insertion of dynamical quarks can 
be handled as well in the large $N$ limit, leading to the celebrate 
't Hooft solution for the meson spectrum \cite{Hooft}. 
On the other hand, in the 
same years, a different prescription for the exchange potential was 
proposed by T.T. Wu in \cite{Wu}. It was derived from analyticity 
requirements and, unexpectedly, it produced a very different 
Bethe-Salpeter equation for the meson wave function; its numerical 
solution \cite{Webbe} exhibits a spectrum that does not resemble 't Hooft's 
one, as confirmed by the analytical computations in \cite{scu}. 
More recently \cite{noi1} it was noticed that Wu's prescription is 
not a mathematical artifact, but it is the the two-dimensional version 
of the four-dimensional Mandelstam-Leibbrandt prescription \cite{WML}, 
that is the {\it only one} reproducing the correct perturbative physics in $4D$ 
\cite{noi3}, when light-cone gauge is adopted. 

The exact resummation of the perturbative series for the Wilson loop 
has been done in \cite{Stau} using the Wu-Mandelstam-Leibbrandt 
(WML) prescription, 
generalizing to all orders the ${\cal O}(g^4)$ computation of 
\cite{noi1,noi2}; it leads, as firstly noticed in \cite{noi1}, 
to a result different from a pure area-law exponentation
(which would be expected from the area-preserving diffeo-invariance of the
theory plus positivity arguments), and, in particular, predicting a different 
value for the string tension. More dramatically, the large-$N$ limit has 
not a confining behaviour, while one easily realizes that, on the plane, the 
theory should be in the strong coupling phase (the plane being thought as 
decompactification of a large sphere).

In \cite{Us} it was shown that the WML computations presented in 
\cite{Stau,noi1,noi2} are indeed ${\it perturbatively}$ correct, in the sense 
that what is missing represents the instanton contribution: the 
WML result for the Wilson loop corresponds {\it exactly} 
to the zero-instanton result of the sphere (that is manifestly gauge 
invariant), when the decompactification limit is taken into account. 
In this perspective the 
absence of confinement at large $N$ is not a mistery, because in the 
weak phase only the zero-instanton sector contributes \cite{Grosso}: it 
seems instead surprising that 't Hooft propagator perturbatively captures 
the (non-perturbative) instantonic physics of the sphere. The gauge 
independence of the perturbative (WML) result has been checked by using 
the Feynman gauge, in which the theory does not look free (the 
self-interaction terms are present there): due to infrared 
singularities the computation has been performed in $D=2+\epsilon$ at 
$O(g^4)$ order, and, apart further terms linked to a possible 
discontinuity at $D=2$ (see ref. \cite{uomo} for a detailed discussion of 
this point), it reproduces the zero-instanton calculus.

If the full WML result has been correctly interpreted, it should be 
related to the local behaviour of the theory: in particular it should be 
derived starting 
from any topology, when the decompactification limit is taken into 
account. To further confirm the results of \cite{Us}, we have decided to 
perform the computation of the zero-instanton contribution to 
Wilson loops on a torus ($T^2$) and to compare it, in the limit of large 
volume, to the WML resummation (we have therefore to consider homologically 
trivial loops on $T^2$ in order to make a comparison). The calculation 
is more difficult than in the sphere case and it presents a certain 
amount of interest by itself. First of all on $T^2$, at variance with the genus 
zero case \cite{Grosso}, no one, at least at our knowledge, has derived the 
explicit form of the partition function and of the Wilson loops in the 
instanton representation. The task is harder from a technical point of 
view, due to the complexity of performing for a Wilson loop 
the Poisson resummation (that is the common 
trick to derive the instanton representation from the 
usual lattice--heat-kernel one). Conceptually we are 
faced with a larger class of classical solutions and non-trivial 
small-coupling singularities due to the non-smooth topology of the 
moduli space of flat connections ${\cal M_F}(T^2, U(N))$.

The paper is organized in the following way. In sect. 2 we rewrite the 
Migdal's partition function for $U(N)$ as a sum of terms localized over the 
classical solutions of Yang-Mills equations. The classical contributions 
appear exactly according to the general classification, presented by Atiyah 
and Bott in \cite{capi}, of the critical points of the Yang-Mills functional 
on Riemann surfaces. We discuss the structure of the quantum 
fluctuations and of the small-couplings singularities, related to 
the non smooth behaviour of ${\cal M_F}(T^2, U(N))$. In sect. 3 we explicitly 
derive the Wilson loop expectation value (for an homologically trivial loop) 
for $U(2)$ and $U(3)$, showing how to perform in concrete the Poisson 
resummation (that is not trivial due to presence of ``dangerous'' 
denominators in the Migdal's sum). The result for the zero-instanton 
contribution coincides, in the decompactification limit, with the WML 
computation. 
In sect. 4 we 
extend the computation to generic $N$, and we find full agreement with 
WML if 
\[
\frac{1}{N+1}{L^{1}_{N}}(x)=\frac{1}{2\pi i}
\int_{{\cal C}_0}\frac{dz}{z^{N+1}}\exp\Biggl[\sum_{k=1}^{+\infty}
L^{0}_{k}(\frac{kx}{N+1})\frac{z^k}{k}\Biggr],
\]
$L^{\alpha}_{N}(x)$ being the generalized Laguerre polynomials (see 
\cite{gradi} for the conventions) and ${\cal C}_0$ a small contour in the 
complex plane surrending the origin: we have not been able to prove 
this identity for generic $N$, but we have checked it  by using MAPPLE 
\footnote{I thank Massimo Pietroni for the help on this point} 
till $N=25$. In sect. 5 we draw our conclusions and discuss possible 
extensions of the work.

After having completed the computations, it has appeared a paper 
by Bill\'o et alt. 
\cite{billo}, in which the relevance of topologically non-trivial 
excitations for $YM_2$ on the torus are discussed and the analogy 
with similar contributions in Matrix String Theory (MST) \cite{DVV} has been 
stressed. Infrared and ultraviolet properties of the partition function 
of MST have also been investigated in \cite{Kosto}: we think that these 
results could be understood as coming from instantons as well, and are 
in some way related to ours.

\section{Partition function and instantons}

We begin by presenting the Migdal's heat-kernel representation for the 
Yang-Mills partition function on $T^2$
\begin{equation}
\label{partition}
{\cal Z}^{(1)}=\sum_{R} \exp\left[-{{g^2 A}\over 2}C_2(R)\right],
\end{equation}
where $A$ is the area of the torus, $g^2$ is the coupling constant 
appearing in the Yang-Mills action and $C_2(R)$ is the value of the 
second Casimir operator in the representation $R$. The sum runs over the 
irreducible representation of the gauge group: in the $U(N)$ case the 
representations $R$ can be labelled by a set of integers 
$n_i=(n_1,...,n_N)$, related to the Young tableaux, obeying the ordering 
$+\infty>n_1>n_2>..>n_N>-\infty$. In terms of $n_i$ we have for the 
second Casimir
\begin{equation}
\label{casimiri}
C_2(R)=C_2(n_1,..,n_N)
=\frac{N}{12}(N^2-1)+\sum_{i=1}^{N}(n_{i}-\frac{N-1}{2})^2.
\end{equation}
The dependence on the product $g^2A$ is peculiar of two dimensional 
Yang-Mills theories, that are invariant under area-preserving 
diffeomorphisms.

Using the permutation symmetry we get
\begin{equation}
\label{partitione}
{\cal Z}^{(1)}=\frac{1}{N!}\sum_{n_1\neq n_2\neq..\neq n_N} 
\exp\left[-{{g^2 A}\over 2}C_2(n_1,..n_N)\right];
\end{equation}
we notice that, at variance with the sphere case, we cannot let two or 
more $n_i$' s be the same. The genus zero partition function  is in fact
\begin{equation}
\label{partitio}
{\cal Z}^{(0)}=\sum_{R}{d_R}^2 \exp\left[-{{g^2 A}\over 2}C_2(R)\right],
\end{equation}
where $d_R$ is the dimension of the representation $R$, that can be 
expressed in terms of the $n_i$'s as 
\begin{equation}
d_R=\Delta(n_1,...,n_N),
\end{equation}
$\Delta$ being the Vandermonde determinant, that possess the permutation 
symmetry and vanishes when two arguments coincides. 
In the following we will see that 
the different structure of the classical solutions on the two topologies 
is reflected by this different behaviour under summation.

The easiest way to obtain the instanton representation is to perform a 
Poisson resummation in eqs.(\ref{partitione},\ref{partitio}) \cite{Poli}: 
this can be done straightforwardly in eq.(\ref{partitio}) 
(see \cite{Grosso} for details), where the sum is extended without problems 
over all the integers, due to the vanishing of the Vandermonde 
determinant, while some care is needed for eq. (\ref{partitione}). 
The partition function can be represented in the following way:
\begin{equation}
{\cal Z}^{(1)}=\frac{1}{N!}\sum_{\{n_i\}=-\infty}^{+\infty}
\sum_{P}(-1)^{P}\int_{0}^{2\pi}
\prod_{i=1}^{N}\frac{d\theta_i}{2\pi}\exp\left[-\sum_{j=1}^{N}\theta_{j}(n_j-
n_{P(j)})\right]\exp\left[-{{g^2 A}\over 2}C_2(n_1,..,n_N)\right],
\label{rapre}
\end{equation}
where no restriction appears on the $n_i$'s. $\sum_{P}$ means the 
sum over all elements of the symmetric group $S_N$, $P(i)$ denotes the 
index $i$ transformed by $P$, while $(-1)^{P}$ is the parity of the 
permutation. 
One recovers the original form eq.~(\ref{partitione}) by simply integrating 
over the angles $\theta_i$ and using the formula
\begin{equation}
\sum_{P}(-1)^{P}\prod_{i=1}^{N}\delta_{n_i,n_{P(i)}}=\det \,
\delta_{n_i,n_j}.
\end{equation}
The basic observation is now that, due to the symmetry of 
$C_2(n_1,...n_N)$, only the conjugacy classes of $S_{N}$ are relevant 
in computing the series: to see this we use the cycle decomposition of 
the elements of $S_N$.

A conjugacy class of $S_N$ is conveniently described by the set of non-negative 
integers $\left\{\nu_i\right\}=(\nu_1,\nu_2,...,\nu_N)$ 
(we follow the description of \cite{libro}) satisfying the constraint
\begin{equation}
\nu_1+2\nu_2+3\nu_3+...+N\nu_N=N.
\end{equation}
Every element belonging to $\{\nu_i\}$ has the same parity and 
can be decomposed, in a standard way, into $\nu_1$ one-cycles, $\nu_2$ 
two-cycles,..., $\nu_N$ $N$-cycles. 
Due to the symmetry of $C_2(n_1,...,n_N)$ all 
the elements of a conjugacy class give the same contribution in 
eq.~(\ref{rapre}), as a simple relabelling of the $n_i$'s and $\theta_i$'s 
is sufficient: only the parity of the class and the 
number of its elements, as function of $\{\nu_i\}$, are therefore relevant 
to the computation of the partition function. It turns out 
that $(-1)^{\sum_{i=even}\nu_{i}}$ is the parity, while 
the number of elements in the conjugacy class $\{\nu_i\}$ is
\begin{equation}
M_{\{\nu_i\}}=\frac{N!}{1^{\nu_1}\nu_1!\,2^{\nu_2}\nu_2!..N^{\nu_N}\nu_N!}.
\end{equation}
The next step is to use the decomposition in cycles 
to perform explicitly the angular integrations: the effect is to express 
the full series as a finite sum of series over a decreasing number of 
integers. One easily realizes that a two-cycle results into the 
identification of two $n_i$'s in the sum, a three-cycle into the 
identification of three $n_i$'s and so on. We end up with
\begin{equation}
{\cal Z}^{(1)}=\frac{{\rm e}^{-\frac{g^2A}{24}N(N^2-1)}}{N!}\sum_{\{\nu_i\}}
\sum_{n_1,..n_\nu=-\infty}^{+\infty}(-1)^{\sum_{i=even}^{}\nu_i} M_{\{\nu_i\}}
\exp\left[ -\frac{g^2A}{2}C_{2}^{\{\nu_i\}}(n_1,..,n_\nu)\right],
\end{equation}
where each conjugacy class has produced a sum over 
$\nu=\nu_1+\nu_2+..+\nu_N$ integers and 
\begin{eqnarray}
C_{2}^{\{\nu_i\}}(n_1,..,n_\nu)&&=
\sum_{i_1=1}^{\nu_1}(n_{i_1}-\frac{N-1}{2})^{2}+
2\sum_{i_2=\nu_1+1}^{\nu_1+\nu_2}(n_{i_2}-\frac{N-1}{2})^{2}\nonumber\\
&&+3\sum_{i_3=\nu_1+\nu_2+1}^{\nu_1+\nu_2+\nu_3}(n_{i_3}-\frac{N-1}{2})^{2}+
....
\end{eqnarray}
Of course if some $\nu_j$ is zero, the integers 
$n_{\nu_1+..\nu_{j-1}+1},..,n_{\nu_1+..\nu_{j-1}+\nu_{j}}$ do not 
appear.
The Poisson resummation is, at this point, almost trivial, being the set 
($n_1,..,n_\nu$) unrestricted: the simple formula 
\begin{equation}
\sum_{n=-\infty}^{+\infty}f(n)=\sum_{m=-\infty}^{+\infty}
\int_{-\infty}^{+\infty}dx\,e^{2\pi imx}f(x),
\end{equation}
requires in our case only gaussian integrations. The final result, 
expressing the original partition function as a sum over ``dual'' integers 
$m_i$'s, is:
\begin{equation}
{\cal Z}^{(1)}=
{\rm e}^{-\frac{g^2A}{24}N(N^2-1)}
\sum_{\{\nu_i\}}\sum_{m_1,..m_\nu=-\infty}^{+\infty}
(-1)^{\Phi_{\{\nu_i\}}(m_1,..,m_\nu)}(\frac{2\pi}{g^2A})^{\frac{\nu}{2}}
Z_{\{\nu_i\}}\exp\left[-S^{\{\nu_i\}}(m_1,..,m_\nu)\right],
\label{closer}
\end{equation}
where
\begin{eqnarray}
\Phi_{\{\nu_i\}}(m_1,..,m_\nu)&&=(-1)^{\sum_{i=even}^{}\nu_i}\exp\left[i\pi(N-1)
(\sum_{j=odd}\sum_{i_{j}}m_{i_{j}})\right],\nonumber\\
Z_{\{\nu_i\}}&&=\frac{\left[1^{\nu_1}2^{\nu_2}3^{\nu_3}...N^{\nu_N}\right]
^{-\frac{3}{2}}}{\nu_1!\nu_2!\nu_3!...\nu_N!},
\end{eqnarray}
and
\begin{eqnarray}
S^{\{\nu_i\}}(m_1,..,m_\nu)&&=\frac{2\pi^2}{g^2A}\sum_{i_1=1}^{\nu_1}m_{i_1}^2+
\frac{\pi^2}{g^2A}\sum_{i_2=\nu_1+1}^{\nu_1+\nu_2}m_{i_2}^2+
\frac{2\pi^2}{3g^2A}\sum_{i_3=\nu_1+\nu_2+1}^{\nu_1+\nu_2+\nu_3}
m_{i_3}^2+....\nonumber\\
&&+\frac{2\pi^2}{Ng^2A}\sum_{i_N=\nu_1+..}^{\nu}m_{i_N}^2,
\label{pleasure}
\end{eqnarray}
and the explicit form of $M_{\{\nu_{i}\}}$ is taken into account. 

These formulae have a nice interpretation in terms of instantons: we 
briefly recall, therefore, the general solution of Yang-Mills equations 
on a compact Riemann surface $\Sigma$ of genus $G$ and with gauge group 
$H$ \cite{capi}. The space of gauge equivalent connections $A$, 
satisfying $D_A\,^*F(A)=0$ ($F$ is the curvature of $A$, $*$ is the 
Hodge operation and $D_A$ is the usual covariant derivative respect 
$A$), can be conveniently described by introducing a suitable central 
extension $\Gamma_R$ of the fundamental group 
$\Pi_1(\Sigma)$ of $\Sigma$ (the central extension is universal and the 
center is extended to $R$). 
The general theorem of Atiyah and Bott states that there is a one 
to one correspendence between the equivalence classes of connections 
which are solutions of the Yang-Mills equations on $\Sigma$ and the 
conjugacy classes of homomorphisms $\rho:\Gamma_R\rightarrow H$. More 
explicitly the connection $A^{(\rho)}$ associated to $\rho$ has curvature 
$F^{(\rho)}=X^{(\rho)}\otimes\omega$, where $\omega$ is the volume form of 
$\Sigma$; $X^{(\rho)}$ is an element of the Lie algebra of $H$ defined by the map 
$d\rho:R\rightarrow {\rm Lie}_H$. As far as we are concerned with the 
evaluation of the Yang-Mills action on the classical solutions, we only have 
to find all the possible $X^{(\rho)}$: this can be easily done for $H=U(N)$. 
The key observation is that, in our case, the homomorphism $\rho$ simply provides 
an unitary representation of $\Gamma_R$. When this representation is 
{\it irreducible}, $X^{(\rho)}$ is central with respect to the action of $H$, 
therefore its eigenvalues are all equal to a certain real number 
$\lambda$: because the Chern class of a $U(N)$ bundle is an integer, 
$\frac{1}{2\pi}\int {\rm Tr}~^*F=k$, we have that the direct computation
\begin{equation}
\frac{V}{2\pi}{\rm Tr}\,X_\rho=k
\label{auto}
\end{equation}
completly determines 
\begin{equation}
\lambda=\frac{2\pi}{V}\frac{k}{N}
\end{equation}
($V$ is the volume of $\Sigma$). 
If $\rho$ is not irreducible we have, in general, that 
$X^{(\rho)}$ is central with respect to a subgroup of $U(N)$ of the type
\begin{equation}
H_X=U(N_1)\otimes U(N_2)\otimes...\otimes U(N_r),
\end{equation}
with $N_1+N_2+..+N_r=N$. This effectively reflects itself into a reduction of 
the original bundle structure ($P(\Sigma,H)$ is a principal $H$-bundle on the
manifold $\Sigma$) 
\begin{equation}
P\left(\Sigma,U(N)\right)\rightarrow \sum_{i=1}^{r}\oplus 
P\left(\Sigma,U(N_i)\right);
\label{riduco}
\end{equation}
the fact that the individual Chern classes must be integers implies, by 
repeating the argument in eq.(\ref{auto}), that $X_\rho$ has eigenvalues 
$\lambda_1,\lambda_2,..,\lambda_r$ with multiplicities $N_1,N_2,..,N_r$:
\begin{equation}
\lambda_i=\frac{2\pi}{V}\frac{k_i}{N_i} 
\end{equation}
($\sum_{i}k_i=k$ is of course the total Chern class). It is possible to 
prove that, fixed the topology of the bundle (that for $U(N)$ is 
determined by a single integer, that is the Chern class $k$), the absolute 
minimum of the Yang-Mills functional is reached when all the eigenvalues 
are equal ($\lambda=\displaystyle{\frac{2\pi}{V}\frac{k}{N}}$ for $U(N)$). 
The action at this minima is
\begin{equation}
S_{{\rm min.}}=\frac{2\pi^2}{g^2V}\frac{k^2}{N}.
\end{equation}
The general Yang-Mills connection, for a $U(N)$ bundle, 
is therefore a direct  sum of Yang-Mills minima for sub-bundles, 
according to the decomposition eq. (\ref{riduco}). The eigenvalues of 
$X_\rho$ can be rational with denominator $N$, when $\rho$ is irreducible, 
while in the opposite situation (if a $\rho$ reducible to $U(1)^N$ exists) 
they are all integers: all the intermediate possibilities according to 
eq.~(\ref{riduco}) could appear. 

The unitary representations of $\Gamma_R$ were 
classified by Narashiman and Seshadri \cite{indi}: they have shown that, 
provided $G>1$, to any pair $(N,k)$ is associated an irriducible 
representation of $\Gamma_R$, producing the Yang-Mills minimum. The 
genus one case is different: in fact for $G=1$, the fundamental group 
$\Pi_1(\Sigma)$ is 
abelian  and so has no irreducible representation for $N>1$. Thus for 
$k=0$ and $N>1$ irreducible representations of $\Gamma_R$ 
do not exist, and this 
corresponds to the well known fact that on $T^2$ every flat connection 
is reducible \cite{Tompo} (we recall that flat connections are 
related the the homomorphisms $\gamma:\Pi_1(\Sigma)\to H$). 
As a consequence one finds that only for $(N,k)$ coprime irreducible 
representations of $\Gamma_R$  do exist. This 
completes our excursion on the well-established (but probably not so 
widely known) subject of the Yang-Mills solutions on Riemann surfaces. 

It is now easy to understand eq.~(\ref{pleasure}) at the light of the 
previous discussions: 

\noindent
$S^{\{\nu_i\}}(m_1,..,m_\nu)$ is the value of the 
Yang-Mills action on a solution determined by $\nu_1$ 
eigenvalues $\lambda_{i_1}=\displaystyle{\frac{2\pi}{A}m_{i_1}}$ 
with multiplicity 1, $\nu_2$ eigenvalues 
$\lambda_{i_2}=\displaystyle{\frac{\pi}{A}m_{i_1}}$ with 
multiplicity 2,..., $\nu_{N}$ eigenvalues 
$\lambda_{i_N}=\displaystyle{\frac{2\pi}{A}m_{i_N}}$ 
with multiplicity $N$, corresponding to the reduction 
$U(1)^{\nu_1}\otimes U(2)^{\nu_2}\otimes...\otimes U(N)^{\nu_N}$. 
The only subtle point in eq.~(\ref{closer}) 
is that not all $S^{\{\nu_i\}}(m_1,..,m_\nu)$ are produced by gauge
inequivalent  solutions: we have to identify, as coming from the same 
instanton, some contributions related to different $\{\nu_i\}$.

To understand this point let us consider the simplest cases, 
$U(2)$ and $U(3)$. The partition function for $U(2)$ is written explicitly
\begin{eqnarray}
{\cal Z}^{(1)}(2)&&=\frac{{\rm e}^{-\frac{g^2A}{4}}}{2}
\Biggl(
\sum_{m_1,m_2=-\infty}^{+\infty}\exp\left[-\frac{2\pi^2}{g^2A}(m_1^2+m_2^2)
\right](-1)^{m_1+m_2}(\frac{2\pi}{g^2A})\nonumber\\
&&+\sum_{m=-\infty}^{+\infty}\exp\left[-\frac{\pi^2}{g^2A}m^2\right]
\frac{(-1)^m}{\sqrt{2}}(\frac{2\pi}{g^2A})^{\frac{1}{2}}\Bigr);
\label{su2}
\end{eqnarray}
the first term corresponds to solutions coming from the reduction 
$U(2)\rightarrow U(1)\otimes U(1)$, while the second one receives 
contributions only from minima (with all eigenvalues equal, proportional 
to $\displaystyle{\frac{m}{2}}$). Actually we have connections in both the sum 
producing the same value of the action (by taking $m_1=m_2=\hat{m}$ in the 
first sum and $m=2\hat{m}$ in the second one, both representing the minimum 
value of the action for the Chern class $k=2\hat{m}$): 
are these connections gauge inequivalent or not? 
The answer is negative because, as we have seen 
before, for $G=1$ only when $N$ and $k$ are coprime the solutions are 
irreducible: the minima, when the Chern class is even, originate, 
therefore, from 
reducible connections (of type $U(1)\otimes U(1)$). The partition 
function is better rewritten as 
\begin{eqnarray}
{\cal Z}^{(1)}(2)&&=\frac{{\rm e}^{-\frac{g^2A}{4}}}{2}\Biggl(
\sum_{m_1,m_2=-\infty}^{+\infty}\exp\left[-\frac{2\pi^2}{g^2A}(m_1^2+m_2^2)
\right]
\left[(-1)^{m_1+m_2}(\frac{2\pi}{g^2A})+\delta_{m_1,m_2}\frac{1}{\sqrt{2}}
(\frac{2\pi}{g^2A})^{\frac{1}{2}}\right]\nonumber\\
&&-\sum_{m=-\infty}^{+\infty}\exp\left[-\frac{\pi^2}{g^2A}(2m+1)^2\right]
\frac{1}{\sqrt{2}}(\frac{2\pi}{g^2A})^{\frac{1}{2}}\Biggr),
\label{su22}
\end{eqnarray}
where every classical contribution appears together the polynomial part 
coming from the quantum fluctuations: incidentally reducible and 
irreducible solutions are disentangled in this way. Repeating the 
exercise for $U(3)$ we get
\begin{eqnarray}
{\cal Z}^{(1)}(3)&&=\frac{{\rm e}^{-g^2A}}{6}
\Biggl(\sum_{\{m_i\}=-\infty}^{+\infty}\exp\left[-
\frac{2\pi^2}{g^2A}(m_1^2+m_2^2+m_3^2)\right]
\Bigl[(\frac{2\pi}{g^2A})^{\frac{3}{2}}
-\delta_{m_2,m_3}\frac{3}{\sqrt{2}}
(\frac{2\pi}{g^2A})\nonumber\\
&&+\frac{2}{\sqrt{3}}(\frac{2\pi}{g^2A})^{\frac{1}{2}}
\delta_{m_1,m_2}\delta_{m_2,m_3}\Bigr]
-\sum_{m_1,m_2=-\infty}^{+\infty}\exp\left[-\frac{2\pi^2}{g^2A}m_1^2
-\frac{\pi^2}{g^2A}(2m_2+1)^2\right]\frac{3}{\sqrt{2}}(\frac{2\pi}{g^2A})
\nonumber\\
&&+\sum_{m=-\infty}^{+\infty}\Bigl(\exp\left[-\frac{\pi^2}{3g^2A}(3m+1)^2\right]
+\exp\left[-\frac{\pi^2}{3g^2A}(3m+2)^2\right]\Bigr)\frac{2}{\sqrt{3}}
(\frac{2\pi}{g^2A})^{\frac{1}{2}}\Biggr).
\label{su3}
\end{eqnarray}
It is now clear that eq.~(\ref{closer}) can be fully rewritten in this 
way, grouping together the classical action for gauge equivalent 
solutions: we prefer to mantain the form of eq.(\ref{closer}), that 
is better for our purpose. 

Some remarks are now in order; first of 
all we notice the difference with the sphere case. There, only instantons 
coming from the $U(1)^N$-reduction were present, and consequently only 
integer numbers labelled the classical solutions; here we have, in 
general, rational numbers associated to instantons. This is the 
geometrical counterpart of the different algebraic form of the sums in 
eqs.~(\ref{partitione},\ref{partitio}), the exclusion of some integers 
in the first one resulting in a richer instantons structure. 
Next we remark that 
the genus one properties provide a simple identification for gauge 
equivalent contributions to the classical action (no irreducible 
representation is avaible for $(N,k)$ not coprime). Third we notice that 
for $g^2A\rightarrow 0$, only the zero-instanton sector survives, as it 
should: let us discuss this limit, that is actually the important one 
for our final computations. Taking all the $m_i$'s to zero in 
eq.(\ref{closer}) we have (we disregard the exponential term that 
simply represents a contribution to the cosmological constant and can be 
eliminated by choosing a suitable renormalization condition \cite{Witte})
\begin{equation}
{\cal Z}^{(1)}_{(0)}=\sum_{\{\nu_i\}}
(-1)^{\sum_{i=even}^{}\nu_i}(\frac{2\pi}{g^2A})^{\frac{\nu}{2}}
\frac{\left[1^{\nu_1}2^{\nu_2}3^{\nu_3}...N^{\nu_N}\right]
^{-\frac{3}{2}}}{\nu_1!\nu_2!\nu_3!...\nu_N!}.
\label{closer2}
\end{equation}
To compute the sum we observe that the constraint 
$\nu_1+2\nu_2+...+N\nu_N=N$ can be implemented by introducing an angular 
variable to obtain
\begin{eqnarray}
{\cal Z}^{(1)}_{(0)}&&=\sum_{\{\nu_i\}=0}^{+\infty}
\int_{0}^{2\pi}\frac{d\theta}{2\pi}
\exp\Biggl[\,i(\nu_1+2\nu_2+...+N\nu_N-N)\theta\Biggr]
(\frac{\pi}{\alpha})^{\frac{\nu_1+\nu_2+..+\nu_N}{2}}\nonumber\\
&&(-1)^{\nu_2+\nu_4+...}\,\,\,\frac{(1^{-\frac{3}{2}})^{\nu_1}}{\nu_1!}
\frac{(2^{-\frac{3}{2}})^{\nu_2}}{\nu_2!}...
\frac{(N^{-\frac{3}{2}})^{\nu_N}}{\nu_N!},
\label{titti}
\end{eqnarray}
where we have defined $\displaystyle{\frac{g^2A}{2}=\alpha}$. 
The sum over $\nu_i$'s is simple, giving
\begin{equation}
{\cal Z}^{(1)}_{(0)}=\int_{0}^{2\pi}\frac{d\theta}{2\pi}
{\rm e}^{-iN\theta}\exp\left[-\sqrt{\frac{\pi}{\alpha}}\sum_{k=1}^{N}
\frac{{\rm e}^{-ik\theta}(-1)^k}{k^{\frac{3}{2}}}\,\right],
\end{equation}
that can be expressed as a contour integral in the complex plane
\begin{equation}
{\cal Z}^{(1)}_{(0)}=\frac{1}{2\pi i}\int_{{\cal C}_0}\frac{dz}{z^{N+1}}
\exp\left[z\Phi(-z;\frac{3}{2};1)\sqrt{\frac{\pi}{\alpha}}\,\,\right],
\end{equation}
where ${\cal C}_0$ rounds the origin anticlockwise, sufficiently close so that 
the function $\Phi(z;s;\mu)$ 
\begin{equation}
\Phi(z;s;\mu)=\sum_{k=0}^{+\infty}\frac{z^k}{(k+\mu)^s};
\end{equation}
is analytic. Expanding the exponential we finally have
\begin{eqnarray}
{\cal Z}^{(1)}_{(0)}&&=\sum_{k=1}^{N}\frac{a_k(N)}{k!}
(\frac{\pi}{\alpha})^{\frac{k}{2}}\nonumber\\
a_k(N)&&=\frac{1}{2\pi i}\int_{{\cal 
C}_0}\frac{dz}{z^{N-k}}\Phi(-z;\frac{3}{2};1)^k.
\label{Singol}
\end{eqnarray}
According to general analysis \cite{Witte} the terms that do 
not vanish exponentially must be interpreted as the contribution of 
the flat connections to the localization formula. 
This can be naively understood observing 
that as $g^2A\rightarrow 0$ the Yang-Mills partition function 
should be equal to the $BF$ one
\begin{eqnarray}
{\cal Z}_{{\rm Y.M}}&&=\int{\cal D}A\exp\left[-\frac{1}{2g^2}\int 
{\rm Tr}\left[F~^*F\right]\right]=\int{\cal D}A{\cal 
D}B\exp\left[\int({\rm Tr}\left[BF\right]+\omega g^2{\rm Tr} 
\left[B^2\right])\right]
\nonumber\\
{\cal Z}_{{\rm BF}}&&=\int{\cal D}A{\cal D}B\exp\left[\,\int{\rm Tr}\left [BF
\right]\right]=
\int{\cal D}A\,\delta(F).
\end{eqnarray}
If the gauge group $H$ 
and the Riemann surface $\Sigma$ 
are such that the space of flat connections is smooth 
and the gauge group acts freely on it, the zero-instanton sector is a 
polynomial in $g^2A$. The zero-order term is the symplectic volume of 
${\cal M_F}(\Sigma,H)$ 
while the coefficients of the other terms have an interpretation as 
intersection numbers on the moduli space. When $H$ and $\Sigma$ are 
such that the space of flat connections is not smooth or the gauge group 
does not act freely on it (as in the case of $U(N)$ on $T^2$) non-analyticity 
appears in the limit $g^2A\rightarrow 0$ (as we have observed in our 
specific case). It could be possible, as suggested in \cite{Witte}, to 
interpretate the singularities in eq.(\ref{Singol}) as coming from the 
singular structure of ${\cal M_F}(\Sigma,H)$, but this is beyond the 
purposes of the present paper. We simply note, that, because we are 
interested to take eventually the decompactification limit on the 
zero-instanton sector, only the term proportional to 
$\displaystyle{(\frac{\pi}{\alpha})^{\frac{1}{2}}}$ will be relevant
\begin{equation}
{\cal Z}^{(1)}_{(0)}\rightarrow 
\frac{(-1)^{N+1}}{N^{\frac{3}{2}}}(\frac{2\pi}{g^2A})^{\frac{1}{2}}
\end{equation}
when $A\rightarrow +\infty$.

\section{Instanton contributions to Wilson loops: the $U(2)$ and $U(3)$ 
case} 

The next step is to derive the instanton representation for an 
homologically trivial, non self-intersecting Wilson loop on $T^2$: we 
want to consider, in particular, the zero-instanton result and eventually 
take the limit of infinite area, in order to make a comparison with 
\cite{Us}. In genus one the general formula \cite{Ruso} gives
\begin{equation}
\label{wilson}
{\cal W}^{(1)}={1\over {\cal Z}^{(1)} N}\sum_{R,S} \frac{d_{R}}{d_{S}}
\exp\left[-{{g^2 A_1}\over 2}C_2(R)-{{g^2 A_2}\over 2}C_2(S)\right]
\int dU {\rm Tr}[U]\chi_{R}(U) \chi_{S}^{\dagger}(U),
\end{equation}
where $A_1+A_2=A$ are the areas singled out by the loop, $A_1$ being the 
simply connected one: the integral in eq.~(\ref{wilson}) is over the 
$U(N)$ group manifold while $\chi_{R,S}(U)$ is the character of the group 
element $U$ in the $R(S)$ representation. We would like to express 
eq.~(\ref{wilson}) through the set of integers defining $R$ and $S$, to 
compute the integral and, at the end, to perform a Poisson resummation. 
The first two steps are easily done, using the explicit formula for the 
characters of $U(N)$ in terms of the $n_i$'s and the symmetry properties 
of the Vandermonde determinants:
\begin{equation}
\label{wilsonp}
{\cal W}^{(1)}=\frac{1}{{\cal Z}^{(1)}N!}
\sum_{n_1\neq n_2\neq..\neq n_N}
\prod_{j=2}^{N}\frac{(n_1-n_j-1)}{(n_1-n_j)}
\exp\left [-\frac{g^2 A}{2}\sum_{i=1}^N (n_i-\frac{N-1}{2})^2
 -g^2 A_1(n_1-\frac{N}{2})\right];
\end{equation}
we have neglected the cosmological constant contribution, that does not 
play any role and actually cancels out in the final result with the same term 
in the partition function. 
Unfortunately the Poisson resummation cannot be performed simply 
following the lines of the previous section  because of the presence of 
the denominators in eq.~(\ref{wilsonp}): when $n_1$ is equal to some 
$n_i$'s the denominator develops a potential singularity, and therefore 
we need a more careful analysis (we cannot extend the sum 
everywhere and then subtract the coincident subsums). Let us consider, 
for the moment, the simplest case, $U(2)$
\begin{equation}
{\cal W}^{(1)}=\frac{1}{2\,{\cal Z}^{(1)}}
\sum_{n_1\neq n_2}\frac{(n_1-n_2-1)}{(n_1-n_2)}
\exp\left [-\frac{g^2 A}{2}(n_1^2+n_2^2)+\frac{g^2 
A}{2}(n_1+n_2-\frac{1}{2})
-g^2A_1n_1\right].
\label{u2}
\end{equation}
The idea is to use a contour representation for this sum: the basic 
identity we need is
\begin{equation}
\sum_{n\neq 
m}\frac{f(m)}{m-n}=\sum_{n=-\infty}^{+\infty}\frac{1}{2i}\int_{\cal 
C}dz\,\cot(\pi z)\frac{f(z)}{z-n}-\sum_{n=-\infty}^{+\infty}f^{\prime}(n),
\label{base}
\end{equation}
where ${\cal C}$ is a contour on the complex plane enclosing the real axis 
anti-clockwise (we assume, of course, that $f(z)$ has a good behaviour at 
$\pm\infty$ near the real axis and has no singularities inside the 
contour). We have therefore two different contributions in 
eq.~(\ref{u2}):
\begin{equation}
{\cal W}^{(1)}=\frac{1}{2\,{\cal Z}^{(1)}}\left[\,B_1+B_2\,\right],
\end{equation}
with
\begin{eqnarray}
B_1&&=\sum_{n=-\infty}^{+\infty}\exp\left[-\alpha (n^2-n+\frac{1}{2})\right]
\left(\frac{1}{2i}\int_{\cal C}dz\,\cot(\pi z)\frac{z-n+1}{z-n}
\exp\left[-\alpha z^2+z(\alpha -2\alpha_1)\right]\right),\\
B_2&&=\sum_{n=-\infty}^{+\infty}\exp\left[-\alpha (n^2-n+\frac{1}{2})\right]
\frac{d}{dz}\Biggl(\,(z-n+1)\exp\left[-\alpha z^2+z(\alpha 
-2\alpha_1)\right]\Biggr)_{z=n},
\end{eqnarray}
being $\displaystyle{\alpha_1=\frac{g^2 A}{2}}$. 
The Poisson resummation can be done without problems on $B_2$, obtaining
\begin{equation}
B_2=(\frac{\pi}{2\alpha})^{{1\over 
2}}\exp\left[-\alpha_1+\frac{\alpha_1^2}{2\alpha}\right]
\sum_{m=-\infty}^{+\infty}\exp\left[-\frac{\pi^2m^2}{2\alpha}+i\pi 
m\frac{\alpha_1}{\alpha}\right](-1)^{m}(1-\alpha_1+\pi i m).
\end{equation}
$B_1$ deserves a deeper study: using the Laurent expansion for 
$\cot(\pi z)$ and explicitly parametrizing ${\cal C}$ with a real 
parameter $t$, one easily gets the Poisson sum:
\begin{eqnarray}
B_1&&=\sum_{m_1=-\infty}^{+\infty}\sum_{m_2=-\infty}^{+\infty}
\int_{-\infty}^{+\infty}dx\,dy\exp\left[ 2\pi i(m_1 x+m_2 y)\right]
\exp\left[-\alpha x^2+\alpha x-\frac{\alpha}{2}\right]\nonumber\\
&&\frac{1}{\pi}\int_{-\infty}^{+\infty}dt (t-x+1){\rm 
Im}\left[\frac{1}{t-y+i\epsilon}\frac{1}{t-x+i\epsilon}\right]
\exp\left[-\alpha t^2+(\alpha-2\alpha_1)t\right].
\end{eqnarray}
One can check, evaluating the imaginary part, that the only non 
vanishing contribution is
\begin{eqnarray}
B_1=\sum_{m_1,m_2=-\infty}^{+\infty}
\int_{-\infty}^{+\infty}dx\,dy\, &&\exp\left[ 2\pi i(m_1 x+m_2 y)
-\alpha(x^2-x+\frac{1}{2})-\alpha(y^2-y)-2\alpha_1 y\right]
\nonumber\\
&&\frac{1}{\pi}(y-x+1)
CPV\left(\displaystyle{\frac{1}{y-x}}\right).
\label{pippo}
\end{eqnarray} 
We can arrive to a compact formula by using the 
Schwinger representation of the denominator in eq.~(\ref{pippo}) and 
integrating on $x$ and $y$:
\begin{eqnarray}
B_1&&=(\frac{\pi}{2\alpha})
\exp\Biggl[-\alpha_1+\frac{\alpha_1^2}{2\alpha}\Biggr]
\sum_{m_1,m_2=-\infty}^{+\infty}
\exp\left[-\frac{\pi^2(m_1^2+m_2^2)}{\alpha}+2i\pi 
m_1\frac{\alpha_1}{\alpha}\right](-1)^{m_1+m_2}\nonumber\\
&&\Biggl[1-\int_{0}^{+\infty}dt\,\exp\left[-\frac{t^2}{2\alpha}\right]
\sin\left[t\left(\frac{\alpha_1}{\alpha}+\frac{i\pi}
{\alpha}(m_2-m_1)\right)\right]\Biggr].
\label{pippo2}
\end{eqnarray}
The final result can be presented as
\begin{eqnarray}
{\cal W}^{(1)}&&=\frac{\exp\Bigl[-\alpha_1\Bigr]}{2{\cal Z}^{(1)}}
\Biggl[\exp\Bigl[\frac{{\alpha_1}^2}{\alpha}\Bigr]
\sum_{m_1,m_2=-\infty}^{+\infty}
\exp\left[-\frac{\pi^2(m_1^2+{m_2}^2)}{\alpha}+2i\pi 
m_1\frac{\alpha_1}{\alpha}\right](\frac{\pi}{\alpha})W(m_1,m_2)\nonumber\\
&&+\exp\left[\frac{{\alpha_1}^2}{2\alpha}\right]
\sum_{m=-\infty}^{+\infty}\exp\left[-\frac{\pi^2m^2}{2\alpha}+i\pi 
m\frac{\alpha_1}{\alpha}\right](\frac{\pi}{2\alpha})^{{1\over 
2}}W(n)\Biggr]
\end{eqnarray}
with
\begin{eqnarray}
W(m_1,m_2)&&=(-1)^{m_1+m_2}
\Biggl[1-\left(\alpha_1+i\pi(m_2-m_1)\right)\,_{1}F_{1}(1,;\frac{3}{2};
-\frac{\left(\alpha_1+i\pi(m_2-m_1)\right)^2}{2\alpha})\Biggr],\nonumber\\
W(m)&&=(-1)^{m+1}(1-\alpha_1-i\pi m),
\label{sdi}
\end{eqnarray}
$\displaystyle{\,_{1}F_{1}(\gamma;\beta;z)}$ 
being a confluent hypergeometric function \cite{gradi}. 
In this form, the result for the Wilson loop closely resembles the structure of 
eq.~(\ref{su2}); we have the classical terms, related to the classical 
action and the holonomy phase evaluated on the instanton solutions, and 
the quantum contributions around them, represented by $W(m_1,m_2),W(m)$. 
Notice that they naturally appear ordered by their singularity degree in 
$\alpha$, a fact that is welcome because we would like to discuss the 
asymptotic behaviour in $\alpha$. The zero-instanton sector gives, 
of course, the non-exponentially suppressed contribution as 
$\alpha\to 0$
\begin{eqnarray}
{\cal W}^{(1)}_0&&=
\frac{\exp\left[\displaystyle{-\frac{g^2A_1}{2}}\right]}
{1-\displaystyle{\frac{1}{\sqrt{2}}(\frac{g^2A}{2\pi})^{\frac{1}{2}}}}
\Biggl[\exp\left[\frac{g^2{A_1}^2}{2A}\right]
\Bigl(1-\frac{g^2A}{2}\,_{1}F_{1}(1;\frac{3}{2};-\frac{g^2A_1^2}{4A})\Bigr)
\nonumber\\
&&-\exp\left[\frac{g^2A_1^2}{4A}\right]\frac{1}{\sqrt{2}}
(\frac{g^2A}{2\pi})^{\frac{1}{2}}(1-\frac{g^2A_1}{2})\Biggr].
\label{sdo}
\end{eqnarray}
Before discussing eq.~(\ref{sdo}), it is better to recall the results of 
\cite{Us}, where the analogous quantity was derived on the genus zero 
(for arbitrary $N$). There we obtained
\begin{equation}
\label{risultato}
{\cal W}^{(0)}_0=\frac{1}{N}\exp\left[-g^2\frac{A_1A_2}{2A}\right]\,
L_{N-1}^1(g^2\frac{A_1A_2}{A}).
\end{equation}
In the decompactification limit $A\to
\infty, A_1$ fixed ($A_1$ and $A_2$ are the areas singled out by the loop 
on the the sphere, $A=A_1+A_2$), the quantity in the equation above 
{\it exactly} coincides, for any value of $N$, with eq.(11) of 
ref.\cite{Stau}, which was derived following completely different 
considerations. We recall indeed that their result was obtained by a full 
resummation at all orders of the perturbative expansion of the Wilson loop
in terms of Yang-Mills propagators in light-cone gauge, 
endowed with the WML prescription. We notice that ${\cal W}^{(0)}_0$ 
does not exhibit the usual area-law exponentiation;
actually, in the large-$N$ limit, exponentiation (and thereby confinement)
is completely lost, as first noticed in \cite{Stau}.
As a matter of fact, from eq.(\ref{risultato}), taking the limit 
$N\to \infty$, we easily get
\begin{equation}
\label{bessel}
\lim_{N\to \infty} {\cal W}_0=\sqrt{\frac{A_1+A_2}{\hat{g}^2 A_1A_2}}
J_1\Bigl(\sqrt{\frac{4\hat{g}^2A_1A_2}{A_1+A_2}}\Bigr)
\end{equation}
with $\hat{g}^2=g^2 N$.
At this stage, however, this is no longer surprising since 
${\cal W}^{(0)}_0$ does not contain any genuine non perturbative
contribution. If on the sphere $S^2$ we
consider the weak coupling phase $\hat{g}^2 A <\pi^2$, instanton
contributions are suppressed. As a matter of fact, eq.(\ref{bessel})
provides the complete Wilson loop expression in the  weak coupling
phase \cite{Kaza,Lupi}. In turn confinement occurs in
the strong coupling phase \cite{Grosso}. 
For any value of $N$ the pure area law exponentiation follows, after
decompactification, from resummation of all instanton sectors, changing 
completely the zero sector behaviour and, in particular, the value of the 
string tension.

Taking into account these considerations, we can now proceed to 
discuss eq.~(\ref{sdo}), that is the exact analogous on $T^2$ of 
eq.~(\ref{risultato}) (in the $N=2$ case). Here the zero-instanton 
contribution is much more complicated, as we could expect due to the 
presence of a non trivial moduli space of flat connections and to an 
intrisic asymmetry between $A_1$ and $A_2$. Moreover no phase transition 
has been found, at large $N$, on the torus, so we do not expect that the 
zero-instanton contribution is related to some regime of the complete 
theory. Nevertheless when $A\rightarrow +\infty$, with $A_1$ fixed, we 
{\it exactly} recover the decompactification limit of 
eq.~(\ref{risultato}) 
\begin{equation}
{\cal W}^{(1)}_0\rightarrow \frac{1}{2}\exp 
\Bigl[-\frac{g^2A_1}{2}\Bigr](1-\frac{g^2A_1}{2}),
\label{risultato2}
\end{equation}
confirming the independence of the zero-instanton contribution, in the 
decompactification limit, from the topology choosen and coinciding with 
the WML computation on the plane. We notice that, looking at 
eq.~(\ref{sdi}), the quantum fluctuation around the irreducible instantons is
\begin{equation}
{\cal W}^{(1)}_{\rm Irr.}=\frac{\exp\Bigl[-\alpha_1\Bigr]}
{2\,{\cal Z}^{(1)}}
\exp\left[\frac{{\alpha_1}^2}{2\alpha}\right]
\sum_{m=-\infty}^{+\infty}\exp\left[-\frac{\pi^2(2m+1)^2}{2\alpha}+2i\pi 
(m+\frac{1}{2})\frac{\alpha_1}{\alpha}\right]W(m),
\end{equation}
that is as well expressed through the same Laguerre polynomial. In other 
words, when the Chern class is odd, the contribution to the Wilson loop 
coming from fluctuations around the minimum of the Yang-Mills action 
seems to mimic the WML result.

Coming back to eq.~(\ref{sdo}), it is possible to recover WML even 
not performing explicitly the decompactification limit: 
taking $g^2A$ and $g^2A_1$ fixed and performing an expansion in 
$\displaystyle{\frac{g^2A_1^2}{A}}$, 
the first term is again eq.~(\ref{risultato2}). We 
don't know the meaning, if any, of this kind of expansion.

The computation for $U(2)$ can be extended to $U(3)$, with an increasing 
complexity in the non-irreducible sector. The expression to be Poisson 
resummed is this time:
\begin{equation}
{\cal W}^{(1)}=\frac{1}{6{\cal Z}^{(1)}}
\sum_{n_1\neq n_2\neq n_3}\left[1+\frac{2}{(n_1-n_2)}+
\frac{1}{(n_1-n_2)(n_1-n_3)}\right]
\exp\left[-\alpha\sum_{n_i=1}^{3}{n_i}^2+\alpha_1n_1\right],
\label{u3}
\end{equation}
and everything goes as in the $U(2)$ case, except that 
the third term needs a bit more work: we have to extract a second 
derivative and the square of a first derivative, but nothing 
conceptually novel happens. We simply obtain a somewhat more complicated 
quantum contribution, involving integrals similar to the previous ones. 
The final expression is  nevertheless cumbersome, reproducing the expected 
expansion in the same instantons of eq.~(\ref{su3}): 
we present only the result for the zero-instanton sector
\begin{eqnarray}
{\cal W}^{(1)}_0&&=
\frac{\exp\Bigl[-\alpha_1\Bigr]}{1-\displaystyle{\frac{3}{\sqrt{2}}}
(\frac{\alpha}{\pi})^{{1\over 2}}+\displaystyle{\frac{2}{\sqrt{3}}}
\frac{\alpha}{\pi}}
\,\Biggl[\exp\Bigl[\frac{{\alpha_1}^2}{3\alpha}\Bigr]
\frac{2}{\sqrt{3}}\frac{\alpha}{\pi}
\Bigl(1-2\alpha_1+\frac{(2\alpha_1)^{2}}{6}\Bigr)\nonumber\\
&&-\exp\Bigr[\,\frac{{\alpha_1}^2}{2\alpha}\,\Bigl]
3\,(\frac{\alpha}{2\pi})^{{1\over 2}}
\left[1-\frac{2}{3}\alpha_1-\frac{4}{3}\alpha_1\,_{1}F_{1}(1;{3\over 2};
-{\alpha^2_1\over 2\alpha})+\frac{4}{3}{\alpha_1}^2\,_{1}F_{1}(1;{3\over 
2};-{\alpha^2_1\over 6\alpha})\right]\nonumber\\
&&+\exp\Bigl[\,\frac{\alpha_1^2}{\alpha}\,\Bigr]\Biggl
(1-2\alpha_1\,_{1}F_{1}(1;{3\over 2};-{{\alpha_1}^2\over 
2\alpha})+{\alpha_1}^2I_1\Biggr)\Biggr]
\label{disper}
\end{eqnarray}
where
\begin{equation}
I_1=\int_0^{+\infty}dt_1\,dt_2\,\exp\Bigl[
-\frac{t_1^2}{2\alpha}-\frac{t_2^2}{2\alpha}-\frac{t_1t_2}{2\alpha}\Bigr]
\sin(t_1\frac{\alpha_1}{\alpha})
\sin(t_2\frac{\alpha_1}{\alpha}).
\end{equation}
In spite of the complicated expression, the limit of infinite area is 
simple
\begin{equation}
{\cal W}^{(1)}_0\rightarrow \frac{1}{3}\exp 
\Bigl[\,-\frac{g^2A_1}{2}\Bigr]\left(1-g^2A_1+\frac{(g^2A_1)^{2}}{6}\right)=
\exp\Bigl[\,-\frac{g^2A_1}{2}\Bigr]\frac{1}{3}L^{1}_{2}(g^2A_1);
\label{risultato3}
\end{equation}
the WML result is again recovered. Explicitly working out 
eq.~(\ref{disper}), one can check that the expansion in 
$\displaystyle{\frac{g^2A_1^2}{A}}$ still reproduces the same Laguerre 
polynomial.

Eq.~(\ref{disper}) strongly suggests that the computation of the 
zero-instanton sector, with $N$ generic, is really involved for the Wilson 
loops: nevertheless, in the decompactification limit, only the term 
proportional to $\displaystyle{({\frac{\pi}{\alpha})^{1\over 2}}}$ 
is relevant. We do not 
try therefore to compute the full contribution, but, in the next 
section, we shall focus our attention only on the leading term.

\section{Instanton contributions to Wilson loops: general relation with 
the perturbative WML result} 

We are interested in computing the zero-instanton contribution to 
eq.~(\ref{wilsonp}) 
for general $N$, in the limit in which $g^2A\to +\infty$. To 
this aim let us introduce a different set of integers and consider the 
group $U(N+1)$,
\begin{eqnarray}
n_1&&\to n,\nonumber\\
n_1-n_{j+1}&&\to -n_j,\,\,\,\,\,\,\,{\rm for }\,\,j=1,...N.
\end{eqnarray}
Eq.~(\ref{wilsonp}) takes then the form
\begin{eqnarray}
{\cal W}^{(1)}&&=\frac{1}{{\cal Z}^{(1)}(N+1)!}\sum_{n_i\neq n_j\neq 0}
\prod_{k=1}^{N}(\frac{n_k+1}{n_k})\exp\left[-\alpha\sum_{i=1}^{N}{n_i}^2\right]
\nonumber\\
&&\sum_{n=-\infty}^{+\infty}\left[-\alpha(N+1)\Bigl(n-\frac{N}{2}\Bigr)^2+
2(n-\frac{N}{2})\Bigl(\alpha_1-\alpha\sum_{i=1}^{N}n_i\Bigr)\right].
\end{eqnarray}
We can Poisson resum over $n$
\begin{eqnarray}
{\cal W}^{(1)}&&=\frac{\exp\left[\,-\alpha_1+\displaystyle{\frac{{\alpha_1}^2}
{(N+1)\alpha}}\right]}{{\cal 
Z}^{(1)}(N+1)!}\Bigl(\frac{\pi}{(N+1)\alpha}\Bigr)^{{1\over 2}}
\sum_{-\infty}^{+\infty}\exp\left[-\frac{\pi^2m^2}{(N+1)\alpha}+\frac{2\pi 
i m}{(N+1)}\frac{\alpha_1}{\alpha}\right](-1)^{mN}\nonumber\\
&&\sum_{n_i\neq n_j \neq 0}\prod_{k=1}^{N}(\frac{n_k}{n_k+1})
\exp\left[-\alpha\left(\sum_{i=1}^{N}{n_i}^{2}+\frac{1}{N+1}
(\sum_{i=1}^{N} 
n_i)^2\right)-\frac{2}{N+1}(\alpha_1-im)\sum_{i=1}^{N}n_i\right].
\label{start}
\end{eqnarray}
In eq.~(\ref{start}) the first line is exactly what we expect for the 
contribution of irriducible instantons, that, when the quantum number is 
taken to zero, is the same, in the decompactification limit, of the 
zero-instanton sector (as the factor 
$\displaystyle{(\frac{\pi}{(N+1)\alpha})^{{1\over 
2}}}$ clearly shows). On the other hand we can argue, from the results 
of the previous section, that when $m$ is prime with respect to $N+1$, 
the full contribution to the Wilson loop is likely to be a Laguerre 
polynomial. The crucial point is therefore to extract, from the very 
complicated function obtained by Poisson resumming over the $n_i$'s, the 
part not involving other instanton numbers neither modifying the singularity 
behaviour in $\alpha$. The basic object we have to study is the sum 
($\xi=2\displaystyle{\frac{\alpha_1-i\pi m}{(N+1)}}$):
\begin{equation}
I(\alpha,\xi)=\sum_{n_i\neq n_j \neq 0}\prod_{k=1}^{N}(\frac{n_k}{n_k+1})
\exp\left[-\alpha\left(\sum_{i=1}^{N}{n_i}^{2}+\frac{1}{N+1}
(\sum_{i=1}^{N} 
n_i)^2\right)-\xi\sum_{i=1}^{N}n_i\right]:
\label{starter}
\end{equation}
what we need is to pass to the ``dual'' integers $m_i$ by using the 
technique developped in the previous section and to evalute the 
contribution at $m_i=0$. The important observation is that, as far as we 
are concerned with this type of computation, the presence of $\alpha$ in 
the exponential can be disregarded: the difference is relevant only in 
the non-leading behaviour as $\alpha\to +\infty$: 
the exact $\alpha$ dependence, for the 
leading term in the zero-instanton sector, is factorized in the 
term $\exp\left[\displaystyle{\frac{{\alpha_1}^2}{(N+1)\alpha}}\right]
\displaystyle{(\frac{\pi}{(n+1)\alpha})^{{1\over 2}}}$ 
appearing in eq.~(\ref{start}). 
With this remark in mind, we can extract the relevant part from the 
series, that we call $I(\xi)$, without worrying about the convergence 
(for which the presence of $\alpha$ is instead essential). 
Eq.~(\ref{starter}) can be rewritten using the conjugacy class formalism 
as
\begin{eqnarray}
I(\xi)\simeq &&\sum_{\{\nu_i\}}
\sum_{n_i\neq 0}\exp\Bigl[ 
i\pi\sum_{i=even}\nu_i\Bigr]M_{\{\nu_i\}}
\prod_{i_1=1}^{\nu_1}\frac{n_{i_1}+1}{n_{i_1}}
\prod_{i_2=\nu_1+1}^{\nu_1+\nu_2}\Bigl(\frac{n_{i_2}+1}{n_{i_2}}\Bigr)
^{2}.....\nonumber\\
&&\exp\left[-\xi\left(\sum_{i_1=1}^{\nu_1}n_{i_1}
+2\sum_{i_2=\nu_1+1}^{\nu_1+\nu_2}n_{i_2}+....+
N\sum_{i_N=\nu_1+..}^{\nu}n_{i_N}\right)\right].
\label{tombola}
\end{eqnarray}
Due to the absence of $\alpha$ we have the miracolous factorization
\begin{equation}
I(\xi)\simeq \sum_{\{\nu_i\}}\exp\Bigl[i\pi\sum_{i=even}\nu_i\Bigr]
M_{\{\nu_i\}}
\prod_{k=1}^{N}\left[\sum_{n\neq 0}(\frac{n+1}{n})^{k}\exp\Bigl[\xi 
kn\Bigr]\right]^{\nu_k}.
\label{miracolo}
\end{equation}
Let us examine the sum
$\displaystyle{
\sum_{n\neq 0}(\frac{n+1}{n})^{k}\exp\Bigl[\xi kn\Bigr]}$: 
it is not convergent, of course (as we said we missed the convergence 
factor, quadratic in the $n_i$'s, in the exponential and proportional to 
$\alpha$), but it captures the relevant behaviour. In fact the zero-term 
in the dual integers coming from the Poisson resummation originates from
\begin{eqnarray}
\sum_{n\neq 0}(\frac{n+1}{n})^{k}\exp\Bigl[\xi kn\Bigr]&&\simeq
\frac{1}{2i}\int_{\cal C}\frac{dz}{z}\cot(\pi z)(\frac{z+1}{z})^k\exp
\Bigr[-\xi z\Bigl]\nonumber\\
&&-\frac{1}{2\pi i}\int_{{\cal C}_0}\frac{dz}{z}\cot(\pi z)
(\frac{z+1}{z})^k\exp\Bigr[-\xi z\Bigl],
\end{eqnarray}
where ${\cal C}$ is the contour considered in eq.~(\ref{base}) while ${\cal 
C}_0$ is a small circle around the origin in the complex plane (we 
remark again that we do 
not mind about the convergence of the first integral; it would be cured 
inserting the $\alpha$ dependence and it would be relevant only for the 
non-leading behaviour in $\alpha$). The $m_i=0$ term comes therefore 
from the second integral and can be nicely evaluated to be:
\begin{equation}
\frac{1}{2\pi i}\int_{{\cal C}_0}\frac{dz}{z}\cot(\pi z)(\frac{z+1}{z})^k\exp
\Bigr[-\xi z\Bigl]=L^0_k(\xi k),
\end{equation}
$L^0_k(x)$ being the usual Laguerre polynomial of degree $k$. In 
eq.~(\ref{miracolo}) we have to consider only terms of the this type, 
obtaining
\begin{equation}
\left[\sum_{n\neq 0}(\frac{n+1}{n})^{k}\exp\Bigl[\xi 
kn\Bigr]\right]^{\nu_k}\simeq 
(-1)^{\nu_k}\left[L^0_k(\xi k)\right]^{\nu_k}.
\end{equation}
The exact leading contribution turns out to be
\begin{equation}
I(\xi)=\sum_{\{\nu_i\}}\exp\Bigl[i\pi\sum_{i=even}\nu_i\Bigr]
M_{\{\nu_i\}}
\prod_{k=1}^{N}\left[L^0_k(\xi k)\right]^{\nu_k}:
\label{miracolo2}
\end{equation}
exploiting the same trick in eq.~(\ref{titti}) we get
\begin{eqnarray}
I(\xi)&&=(-1)^N\sum_{\nu_k=0}^{+\infty}\frac{1}{2\pi}\int_{0}^{2\pi}d\theta
e^{-iN\theta}
\prod_{k=1}^{N}\left[\displaystyle{\frac{L^0_k(\xi k)}{k}}\right]^{\nu_k}
\displaystyle{\frac{\exp \Bigl[i\theta k 
\nu_k\Bigr]}{\nu_k!}}\nonumber\\
&&=(-1)^{N}N!
\int_{{\cal C}_0}\frac{dz}{z^{N+1}}\exp\left[\sum_{k=1}^{N}L_k^0(\xi 
k)\displaystyle{\frac{z^k}{k}}\right].
\label{miracolo3}
\end{eqnarray}
We have the elements to extract the zero-instanton 
contribution in the limit $\alpha\to +\infty$: for the partition 
function
\begin{equation}
{\cal Z}^{(1)}_{(0)}\to
(-1)^{N}\frac{1}{N+1}\left(\frac{\pi}{(N+1)\alpha}\right)^{\frac{1}{2}},
\end{equation}
then it follows (for $m=0$ we have 
$\displaystyle{\xi=\frac{2\alpha_1}{N+1}}$)
\begin{equation}
{\cal W}^{(1)}_0\simeq
\exp\left[\,-\alpha_1+\frac{{\alpha_1}^2}{(N+1)\alpha}\right]
\frac{1}{2\pi i}\int_{{\cal C}_0}
\frac{dz}{z^{N+1}}\exp\left[\sum_{k=1}^{+\infty}L_k^0
\displaystyle{(\frac{2\alpha_1}{N+1}k)}
\displaystyle{\frac{z^k}{k}}\right];
\end{equation}
We have extended the sum in the exponential of eq.~(\ref{miracolo3}) to 
$+\infty$, taking the contour ${\cal C}_0$ sufficiently close to the 
origin. In the decompactification limit we end up with
\begin{equation}
{\cal W}^{(1)}_0=\exp\left[-\frac{g^2A_1}{2}\right]
\frac{1}{2\pi i}\int_{{\cal C}_0}
\frac{dz}{z^{N+1}}\exp\left[\sum_{k=1}^{+\infty}L_k^0\left(
\displaystyle{\frac{g^2A_1}{N+1}k}\right)
\displaystyle{\frac{z^k}{k}}\right],
\end{equation}
that is the desired result: the calculation of an homologically trivial 
Wilson loop on the torus, taking into account only the zero-instanton 
sector and then decompactifying the torus itself. We see that if the 
formula
\begin{equation}
\frac{1}{N+1}{L^{1}_{N}}(x)=\frac{1}{2\pi i}
\int_{{\cal C}_0}\frac{dz}{z^{N+1}}\exp\Biggl[\sum_{k=1}^{+\infty}
L^{0}_{k}(\frac{kx}{N+1})\frac{z^k}{k}\Biggr],
\label{Mira}
\end{equation}
is true, we have full agreement with the WML computation on the plane 
and with the same calculation done from the sphere. We have not been 
able to prove eq.~(\ref{Mira}) for general $N$, but we have used MAPPLE 
to perform the analytical check (the integral is simply the 
$N$-derivative of the exponential respect to $z$, evaluated in $z=0$) 
till $N=25$. The general proof should not be trivial, because a 
non-linearity, quickly increasing with $N$, involves different types of 
Laguerre polynomials. If our conjecture is true (and we think that the 
MAPPLE result and the perfect agreement with WML are not accidental), we have 
proven that on genus one, when only the 
zero-instanton sector is taken into account, 
the Wilson loop averages coincide with the 
perturbative WML resummation on the plane (in the decompactification 
limit). This suggests that the WML prescription is truly related to the 
local behaviour of ${\rm YM}_2$, considering only the trivial solution in a 
path-integral expansion. We notice that while on the sphere flat 
connections do not exist (in the sense that they are all gauge 
equivalent to the trivial one) on the torus, although reducible, they 
are definitively there, complicating the expansion around $F=0$: 
nevertheless the leading term as $A\to +\infty$ still reproduces WML. 
This fact has probably some meanings in relation to the structure of 
${\cal M_F}(T^2,U(N))$, but we do not enter into the question. Moreover we 
see that all the irriducible connections (minima of the 
Yang-Mills action when the Chern class $k$ is prime with respect to $N$) 
contribute to 
the Wilson loop with $L^1_N(g^2A_1-i\pi m)$, as it is evident from 
eq.~(\ref{miracolo3}). The fluctuation around these particular instantons (that 
are the absolute minima inside the topological class $k$ coprime to $N$) 
exactly produces the Laguerre polynomial, a fact that probably deserves a 
deeper interpretation. Actually we notice that on the sphere too 
there were minima contributing in a similar way: it is not difficult to 
show that when the Chern class is $k=mN$ the minimum of the action is 
obtained for $m_i=m$ and in the expression for $W(m_1,..,m_N)$ the usual 
Laguerre polynomial appears.

\section{Conclusions}
The instanton contributions to the $U(N)$ Yang-Mills partition function and to 
homologically trivial Wilson loops on $T^2$ have been carefully 
examined. We have derived the instanton representation for the partition 
function and identified the contributions coming from the classical 
solutions of Yang-Mills equations on genus one: they appear according to 
the general classification of Atiyah and Bott \cite{capi}, allowing for 
rational values of the instanton numbers, at variance with the sphere 
case, where only integer Dirac monopoles were present. We discussed the 
singularities of the zero-instanton sector and their relevance in the 
decompactification limit. Then we developed a technique to perform the 
Poisson resummation for homologically trivial Wilson loops: at variance 
with the sphere case we are faced with potential singularities in 
passing from the Young tableaux indices $n_i$'s (coming from the Migdal's 
representation of the Wilson loop averages) to the ``dual'' integers 
$m_i$'s (labelling the instanton solutions). We presented the full 
result in the simplest cases, namely $U(2)$ and $U(3)$, that are 
nevertheless quite complicated. The Wilson loop appears as a sum over 
classical contributions (coming from the Yang-Mills solutions) modified 
by finite quantum corrections. The zero-instanton part is particulary 
interesting and can be expressed through confluent hypergeometric 
functions: in the large area limit, it reproduces the WML perturbative 
resummation. We have extended this computation to general $N$, finding 
agreement with the sphere and the plane results if a particular 
mathematical identity, eq.~(\ref{Mira}), holds: from a mathematical point 
of view, it should be non-trivial to prove it in general (we 
checked it analytically till $N=25$ by using MAPPLE).

Our computation on $T^2$ has therefore confirmed the conclusion in 
\cite{Us}, namely that 
there is no contradiction between
the use of the WML prescription in the light-cone propagator and the pure
area law exponentiation; this prescription is correct, but the ensuing
perturbative calculation can only provide us with the expression 
for ${\cal W}_0$, the result coming from the decompactification of the 
sphere and of the torus, when instantons are disregarded. 
The paradox of ref.\cite{Stau} is solved
by recognizing that they did not take into account the genuine ${\cal
O}(\exp(-\frac{1}{g^2}))$ non perturbative quantities. As a matter of 
fact, the Migdal's formula for the Wilson loop eq.~(\ref{wilson}) 
can be understood as a full summation of 
perturbative and non-perturbative contributions: in the limit $A\to 
+\infty$, $A_1$ finite, it simply reproduces the pure area law. 
We notice that, 
from the torus point of view, even the full contribution of the 
flat connections must be considered to recover the pure area result.

What might instead be surprising in this context is the fact that, using the 
istantaneous 't Hooft-CPV potential and just
resumming at all orders the
related perturbative series, one still ends up with the
correct pure area exponentiation. This  feature is likely
to be linked to some peculiar properties of the light-front vacuum
(we remind the reader that the light-cone CPV prescription follows
from canonical light-front quantization \cite{noi3}; still
we know it is perturbatively unacceptable in higher dimensions \cite{Cappe} 
and cannot be smoothly continued to any Euclidean formulation).

From another point of view, we find interesting the possibility to 
relate the zero-instanton sector to the structure of ${\cal M_F}(\Sigma, 
U(N))$, and, more generally, to explore homologically non trivial loops 
in this context. On the other hand we think that 
the relation between our results 
(that are obtained in the spirit of the semiclassical expansion, 
according to the localization principle) and the recent computations of 
\cite{billo,Kosto}, where topological sectors appeared to play a crucial 
role, deserves further studies. A first step in this direction 
should be, for example, to understand how the instanton solutions are 
connected to the topological obstructions to smoothly diagonalize 
two-dimensional Lie-valued maps \cite{Blau2}. Investigations in these 
directions are in progress.

\vskip 1.0truecm
${\bf Acknowledgement}$

It is a pleasure to thank Prof. A. Bassetto and D. Seminara for a 
careful reading of the manuscript. A special thank goes to Massimo 
Pietroni for his help in checking eq.~(\ref{Mira}).

\vfill\eject
\end{document}